\documentclass[twocolumn,english]{revtex4-1}
\usepackage[T1]{fontenc}
\usepackage[latin9]{inputenc}
\usepackage{amsmath}
\usepackage{amssymb}
\usepackage{graphicx}
\usepackage{babel}
\begin{document}

\title{TASEP Speed Process: An Effective Medium Approach}

\author{Aanjaneya Kumar and Deepak Dhar}

\affiliation{Department of Physics, Indian Institute of Science Education and
Research Pune}
\begin{abstract}
We discuss the approximate phenomenological description of the motion of a single second-class particle in a two-species totally asymmetric simple exclusion process (TASEP) on a 1D lattice. Initially, the second class particle is located at the origin and to its left, all sites are occupied with first class particles while to its right, all sites are vacant. Ferrari and Kipnis proved that in any particular realization, the average velocity of the second class particle tends to a constant, but this mean value has a wide variation in different histories. We discuss this phenomena, here called the \emph{TASEP Speed Process}, in an approximate effective medium description, in which the second class particle moves in a random background of the space-time dependent average density of the first class particles. We do this in three different approximations of increasing accuracy, treating the motion of the second-class particle first as a simple biassed random walk in a continuum Langevin equation, then as a biased Markovian random walk with space and time dependent jump rates, and finally as a Non-Markovian biassed walk with a non-exponential distribution of waiting times between jumps. We find that, when the displacement at time $T$ is $x_0$, the conditional expectation of displacement, at time $zT$ ($z>1$) is $zx_0$, and the variance of the displacement only varies as $z(z-1)T$. We extend this approach to describe the trajectories of a tagged particle in the case of a \emph{finite} lattice, where there are $L$ classes of particles on an $L$-site line, initially placed in the order of increasing class number. Lastly, we discuss a variant of the problem in which the exchanges between adjacent particles happened at rates proportional to the difference in their labels.
\end{abstract}
\maketitle

\section{Introduction}

There has been a lot of interest in understanding exclusion processes
on a line as a simple model of stochastic evolution in systems of
interacting particles $[1]$. These are good models of many physical
systems, such as traffic on highways $[2]$, transport in narrow channels
$[3]$ and motion of motor proteins on microtubules $[4]$. Many exact
results are known for the simple exclusion process on a line $[5]$.
Several variants of the exclusion process have been studied including
multi-species exclusion processes and the partially and totally asymmetric
exclusion processes (ASEP and TASEP) $[1, 5-8]$.

If we want to study the trajectory of individual particles in an assembly
of interacting particles, one often adopts a self-consistent mean-field
kind of approximation, in which the motion of the particle occurs
in an effective field provided by the others. The best known example
being Brownian motion $[9, 10]$, that was first studied to describe
the motion of pollen grains in a liquid. Other examples of self-consistent
treatments include the Hartree-Fock theory of electronic structure
of atoms $[11, 12]$, and the motion of ions in plasmas in the Vlasov
approximation $[13]$. 

In this paper, we will discuss this general approach, called the \emph{effective
medium approach} here, in the specific setting of a two-species totally
asymmetric exclusion process. We will consider a system of hard-core
particles on a 1-dimensional lattice, with two classes of particles.
We will consider the evolution from the special initial condition,
where there is only one second class particle at the origin, and all
sites sites to the left are occupied by first class particles, and
all sites to the right of the origin are vacant. The dynamics follows
continuous-time Markovian evolution where each first class particle
exchanges position with a second-class particle or vacancy to its
right with rate 1. The second class particle can jump to the left,
if forced by a first class particle moving from its left, or jump
one space to an empty site on its right, with rate 1. 

For this problem, Ferrari and Kipnis found a rather surprising observation
$[14]$. In their own words, ``a second class particle initially
added at the origin chooses randomly one of the characteristics with
the uniform law on the directions and then moves at constant speed
along the chosen one.'' This is a remarkable property as the system
undergoes Markovian evolution, and has no memory. It happens, because
if the second-class particle initially, by chance, gets a large positive
displacement, in subsequent times it encounters a smaller density
of other particles, and hence also moves faster at later times. This
is an example of persistence, where time average of one evolution
history is very different from ensemble average over all histories
of evolution. 

While the authors proved this result, they did not discuss how big
are the fluctuations in the velocity, and how they decrease with time.
In this paper, we will describe this process in a simple Langevin
description $[15]$, that also allows us to estimate how the fluctuations
in the average speed decrease with time. 

We will show that, when the displacement of the second class particle at time $T$ is $x_0$, the conditional expectation of displacement, at time $zT$ ($z>1$) is $zx_0$, and the variance of the displacement only varies as $z(z-1)T$. Thus the fluctuations, for fixed $z$, increase as  $\sqrt{T}$.
Equivalently, we find that if $v^*$ is the asymptotic value of velocity of the second class particle, for large z, $(v^*- x_0/T)$ has a typical spread of $\frac{1}{\sqrt{T}}$
which goes to $0$ as $T$ increases.

The plan of this paper is as follows: in Section II, we define the
model precisely. In section III, we discuss the description of the
trajectory of the second-class particle in a Langevin equation description.
We use this to determine the variance of the particle position at
time $zT$, given the position at time $T$. In Section IV, we discuss
different approximations of increasing accuracy describing the trajectory
as a biased random walk on the integer lattice. We then use this approach
to study the mean trajectories of a second class particle when the
lattice is finite in Section V. We discuss the interesting end-effects
that occur in this finite lattice version, explain it using our effective
medium approach and outline an interesting new direction that arises
through a simple modification in evolution rules. Section VI contains
a summary and concluding remarks. 

\section{Definition of the Process}

We consider a two species TASEP with initial conditions such that
a single second class particle is located at the origin of the lattice.
To the left of the second class particle, each lattice site is occupied
by a first class particle and to its right, each site is vacant. We
will denote a first class particle by 1, a second class particle by
2 and a vacant site by 0. The allowed nearest neighbor transitions
in this process are:

\[ 10~~\rightarrow 01; ~~20 ~~\rightarrow 02; ~~12 \rightarrow 21 \]
We assume that all these transitions occur with rate $1$.

We wish to understand the dynamics of the second class particle in
this process. Ferrari and Kipnis proved $[14]$ that the position
of the second class particle $X(t)$ at time $t$ follows:

\[
\lim_{t\rightarrow\infty}\frac{X(t)}{t}=U
\]
where $U$ is a uniform random variable on $[-1,1]$. 

We will call the process in which the velocity of the second class
particle tends to a random number distributed uniformly between $[-1,1]$
as \emph{TASEP Speed Process} (TSP) and will provide a simple explanation
of this remarkable phenomena. The name TSP earlier has been used in
the study of joint distribution of the velocities of different particles
in a multi-species version of this process $[16]$. 

\begin{figure}[h]
\includegraphics[scale=0.35]{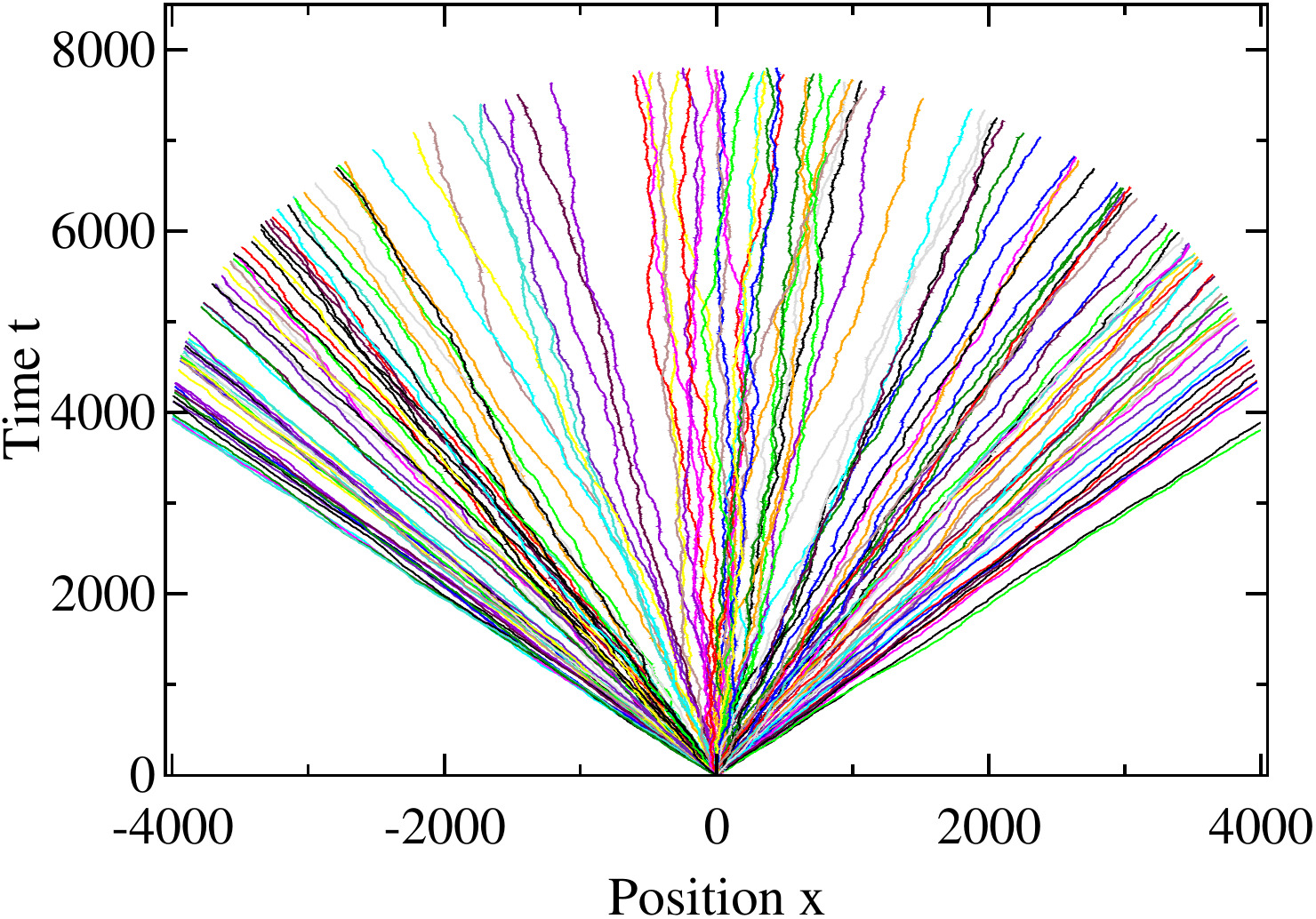}

\caption{Trajectories of second class particle in the TASEP Speed Process.
150 different trajectories consisting of 4000 steps taken by the second
class particle have been plotted. }
\end{figure}

\section{Langevin Description}

We aim to understand the motion of the second class particle, when
all the sites to its left are occupied by first class particles and
all sites to its right are vacant, using a simple approximation by
breaking this problem into two steps:
\begin{enumerate}
\item We first discuss how the mean density $\rho(x,t)$ of first class
particles evolves in space and time, in the absence of the second-class
particle.
\item Then we try to describe the motion of the second class particle moving
as a random walk in a space-time-dependent background field $\rho(x,t)$
. 
\end{enumerate}
We show that this simple description captures essential features of
TSP and allows for further analysis.

Let $x(t)$ denote the position of the second class particle at time $t$. We want to discuss the stochastic properties of this trajectory, by integrating out all the first-class particles. The hydrodynamics of TASEP was first studied by Rost $[17]$. The coarse-grained evolution of the sea of first class particles in terms of particle density $\rho=\rho(x,t)$ can be described by the partial differential equation:
\begin{equation}
\frac{\partial\rho}{\partial t}+(1-2\rho)\frac{\partial\rho}{\partial x}=0
\end{equation}
with
\[ \rho(x,t=0) = \theta(-x), \] where $\theta(x)$ is the step function, which is $0$ for $x<0$, and $1$, for $x >0$. The solution of this partial differential equation is obtained to be:
\begin{eqnarray} \rho(x,t)=1\quad & for\quad & x<-t\\ \rho(x,t)=0\quad & for\quad & x>t\\ \rho(x,t)=\frac{1}{2}(1-\frac{x}{t})\quad & for\quad & -t\leq x\leq t \end{eqnarray}
The motion of the second class particle is described by  a stochastic differential equation 
\begin{equation}
\frac{dx}{dt}=     \bar{V}(\rho(x,t)) + \eta(t) 
\end{equation}
where $\bar{V}$  equals the mean velocity of the particle, and $\eta(t)$  takes into account all the fluctuations away from the mean. By definition, $<\eta(t)> =0$.  $\bar{V}$ is an externally prescribed function of $\rho$ in Eq(5). In our problem,  $\bar{V}=1-2\rho$,  where rho is given by Eq(4). Hence we write
\begin{equation}
\bar{V}(\rho(x,t))= 1- 2 \rho(x,t)
\end{equation}
Substituting the value of $\rho(x,t)$ from above:
\begin{equation}
\frac{dx}{dt}=\frac{x}{t}+\eta(t)
\end{equation}
This is a linear differential equation, and may be solved by using an integrating factor. Equivalently, we make a change of variables to $v(t) = x(t)/t$, the mean velocity of the particle. This satisfies the simpler equation
\begin{equation}
\frac{dv(t)}{dt}=\frac{\eta(t)}{t} 
\end{equation}
This is easily solved to give 
\begin{equation}
v(zT)-v(T)=\intop_{T}^{zT}\frac{\eta(t')}{t'}dt' 
\end{equation}
As $\eta$, by definition has zero mean and $z$ is a real number greater than 1. This gives $<v(zT)>=<v(T)>$. 
\\We can also determine the variance of $v(T)$: 
\begin{equation}
\left\langle \left(v(zT)-v(T)\right)^{2}\right\rangle \:=\intop_{T}^{zT}\intop_{T}^{zT}\frac{<\eta(t')\eta(t'')>}{t't''}dt'dt'' 
\end{equation}
We expect correlation function <$\eta(t) \eta(t')>$ to be short-ranged. It was shown for TASEP in $[18]$ that correlations $< \rho (x,t) \rho(x',t')> $ are exponentially decreasing in time $|t-t'|$, unless $x$ and $x'$ are such that $ x -x' = u( t -t')$, where $u$ is the mean velocity of the flow. In our case, we easily see that while the second-class particle sees a constant density,  the mean velocity of first class particles is $1 -\rho$, and of second class particle is $ 1 -2 \rho$, and they are not equal. So, in general, the correlation function is short-ranged, and if $D = \int_{-\infty}^{+\infty} d\tau  < \eta (t) \eta(t +\tau)>$, we may write $<\eta(t) \eta(t')>=D\delta(t-t')$), which gives
\begin{equation}
<[v(zT) - v(T)]^2> = \frac{D(z-1)}{zT}
\end{equation}
which goes to $0$ as $T$ increases. This shows that the velocity of the second-class particle does get fixed at large $T$. 

\begin{figure}[h]
\includegraphics[scale=0.21]{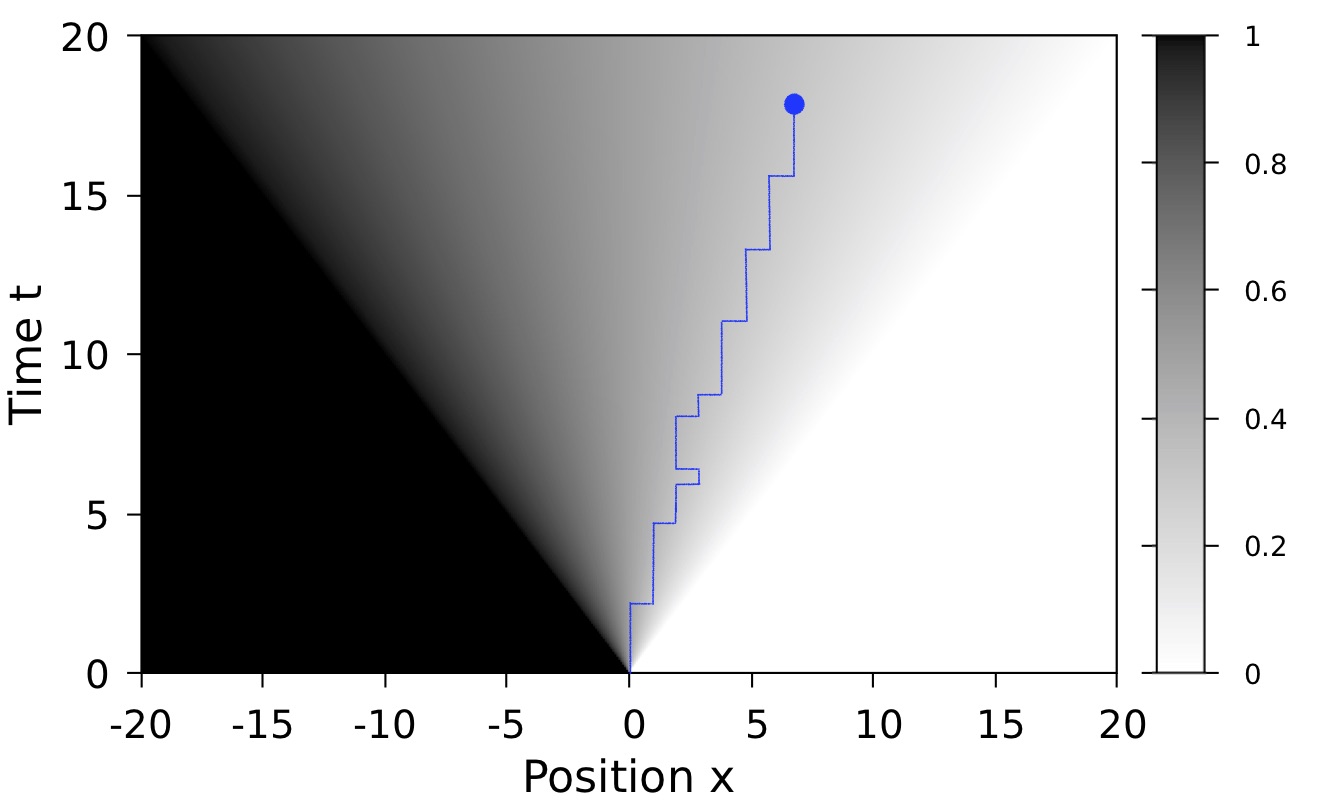}\caption{A schematic representing the motion of the second class particle (blue)
in the space-time dependent background density plotted as a heat map. }
\end{figure}

\section{Approximate random walk descriptions of the trajectory}

While the Langevin description correctly describes the long-time behavior
of trajectory correctly, the actual walk occurs on a discrete lattice,
and a more accurate description would be as a random walk on a line
in continuous time. This we will develop now.

\subsection{Simple Biased Random Walk}

In the spirit of the discussion above,  consider motion of a second class particle in uniform density $\rho$. The trajectory then has mean velocity $U=1-2\rho$, and its time evolution for times $t>>1$ can be discussed as a  simple random walk. It is known that if there is no second class particle, then in the steady state, occupation numbers of TASEP have a product measure.  Then, in the steady state of TASEP, with a fixed density $\rho$,  if we assume that we place a second class particle, with prob.  $(1 -\rho)$, its site on the right will be empty, and then it jumps with rate 1. Similarly, with probability $\rho$, the site on its left will be occupied and it will overtake the second class particle with rate 1. 
So, we conclude that trajectory of particle is a biased random walk. On a background density $\rho(x,t)$, the second class particle jumps to the left with rate $\rho(x,t)$, and to the right with rate $1 -\rho(x,t)$.  As a check, the mean velocity is $U=1 - 2\rho(x,t)$,  which agrees with the exact asymptotic value of velocity $[19]$.  
We have simulated this walk on the background $\rho(x,t)$ given by eq. 2-4. The results are shown in Figure 3. We also compare with the the simulation of the original process  (Figure 1). We see that  while we do get trajectories with velocity fixation, and the velocity $U$ is uniformly distributed in the interval $[-1,1]$, the time taken by the walker to take 4000 steps is roughly the same while, the time taken in TSP shows a clear $\rho$ dependence.

\begin{figure}[h]
\includegraphics[scale=0.35]{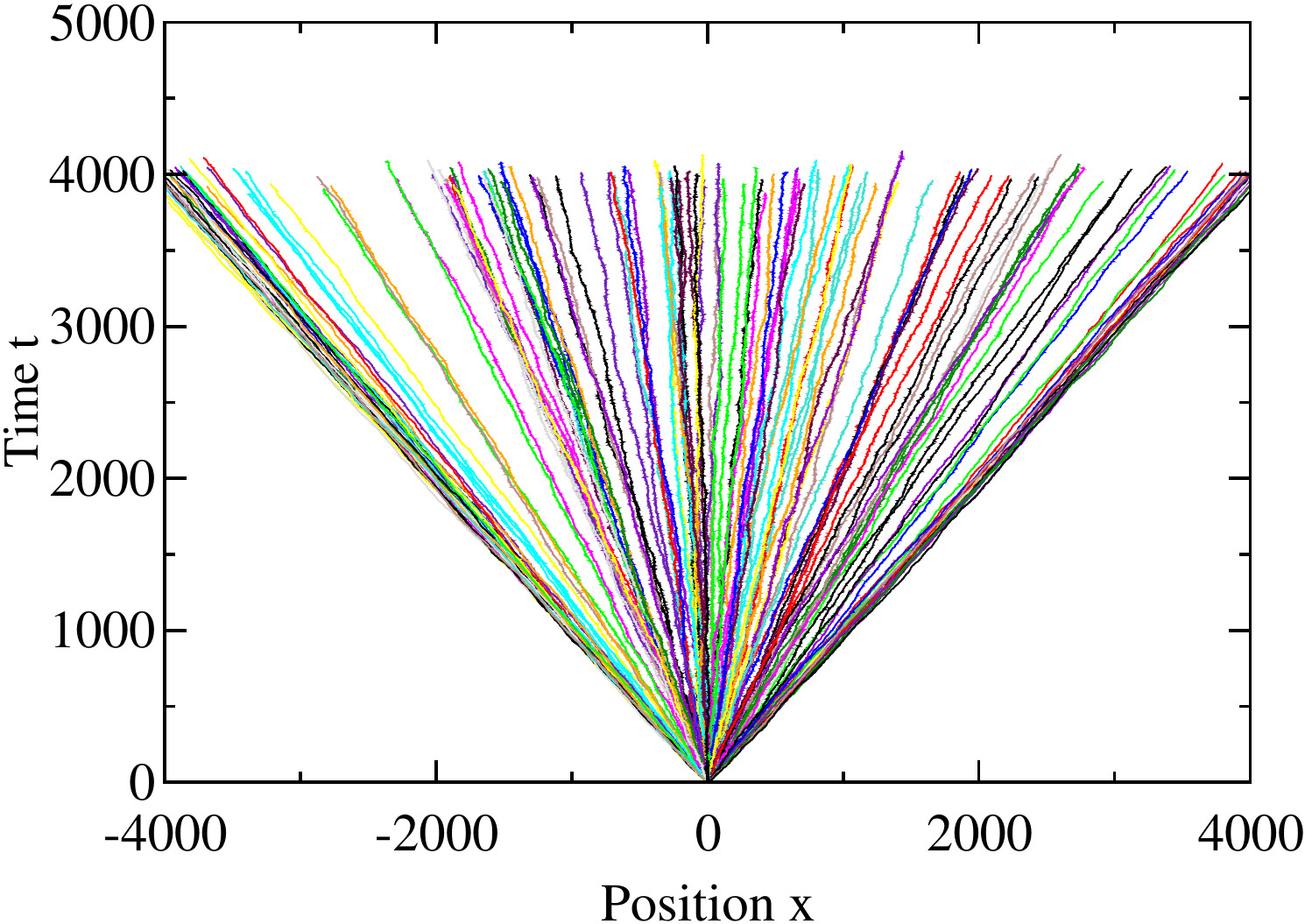}

\caption{Trajectories of a CTRW with rate $\rho(x,t)$ of jumping to the left
and $1-\rho(x,t)$ of jumping to the right. 150 different trajectories
consisting of 4000 jumps made by the random walker have been plotted.
Notice that the time taken by the walker to take 4000 jumps is roughly
the same in each trajectory which is not the case in TSP.}
\end{figure}

\subsection{Markovian Continuous Time Random Walk}

This shows that the rates of left and right jumps in our simple approximate model do not correctly describe the  trajectories of the original problem.  The difference occurs because the average density of first class particles near the second-class particle is not the same as in the bulk, away from the second class particle. Hence our approximation of using the steady state measure of TASEP to calculate jump rates in the problem with a second class particle is present not adequate. The  average density profile near a second class particle in the steady state has been calculated in $[20]$ using the matrix product ansatz. It was shown that the second class particle is attracted to regions with a positive density gradient. More precisely, it was found that for a second class particle on a ring with density $\rho$ of first class particle, in the steady state, the mean density on the site to the right is $2\rho - \rho^2$ and on the site to the left is $\rho^2$. This implies the probability of the site to the right being empty is $(1 - \rho)^2$.
If we use a continuous time random walk model with jump rates $(1 - \rho)^2$ to the right and $\rho^2$ to the left we still get mean velocity = $1 - 2 \rho$. But now the agreement with the simulations is much better as seen in Figure 4.

\begin{figure}[h]
\includegraphics[scale=0.35]{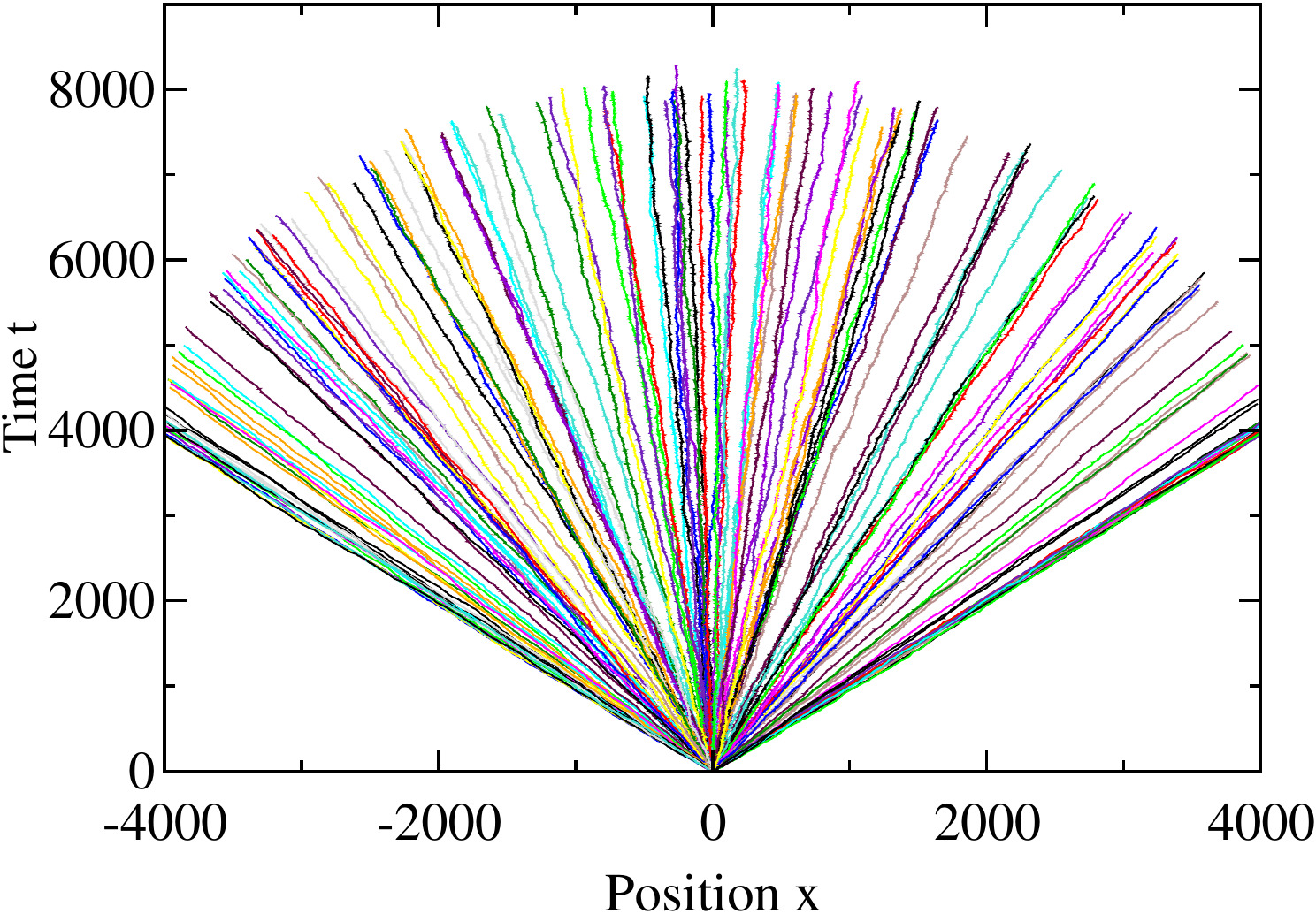}\caption{Trajectories of second class particle in the CTRW model with jump
rates $\left(1-\rho(x,t)\right)^{2}$ to the right and $\left(\rho(x,t)\right)^{2}$
to the left. 150 different trajectories consisting of 4000 steps taken
by the second class particle have been plotted. }
\end{figure}

\subsection{Continuous Time Random Walk with Waiting Time Distributions}

However, our description is still not sufficiently accurate. If we are given a single long trajectory of the second-class particle, with mean velocity $U$ in the original process between times $T$ and $nT$, for $T>>1$, and also one generated using the Markovian jump rates descibed above, can one distinguish between them? The answer is yes. Clearly, in the Markovian approximation, the waiting times between successive jumps are independent random variables, with a distribution that is a simple exponential. 
One can easily verify that in the original process the waiting time intervals do not have an exponential distribution. This comes from the fact that since occupancy of neighbors by first class particles have non-trivial correlations in time, so the probabilities of jump in nearby time intervals $[t, t +\Delta t_1]$ and $ [t +\Delta t_1, t +\Delta t_1 + \Delta t_2]$ are not uncorrelated. 

The trajectories in the TASEP Speed Process show a nearly exponential distribution of waiting times only for velocities close to $1$ and $-1$ whereas a clear departure from the exponential distributiom is observed for smaller velocities. Figure 5 shows the distribution of waiting times for the TSP and the CTRW model with jump rates. The histograms were plotted for two trajectories, consisting of $8000$ steps taken by the second class particle, of velocity (a) 1 and (b) 0.2. It is clearly seen that an even more accurate modelling of the trajectory will be as a random walk which involve a continuous time non-Markovian walk with a prescribed distribution of residence times $f(\tau)$, with probability to jump left or right given by $p(\tau) $ and $1 -p(\tau)$. The calculation of the exact functions $f(\tau)$ and $p(\tau)$ is rather difficult, and will not be attempted here. We can take these to be approximately determined from simulations. 

Of course, even this modelling of the trajectory as a continuous time
random with a distribution of waiting times is approximate. In the
original process, the waiting times between successive jumps are only
approximately uncorrelated. But going beyond this description falls
outside our aim of providing a simple approximate description of the
trajectories. 

\begin{figure}[h]
\includegraphics[scale=0.45]{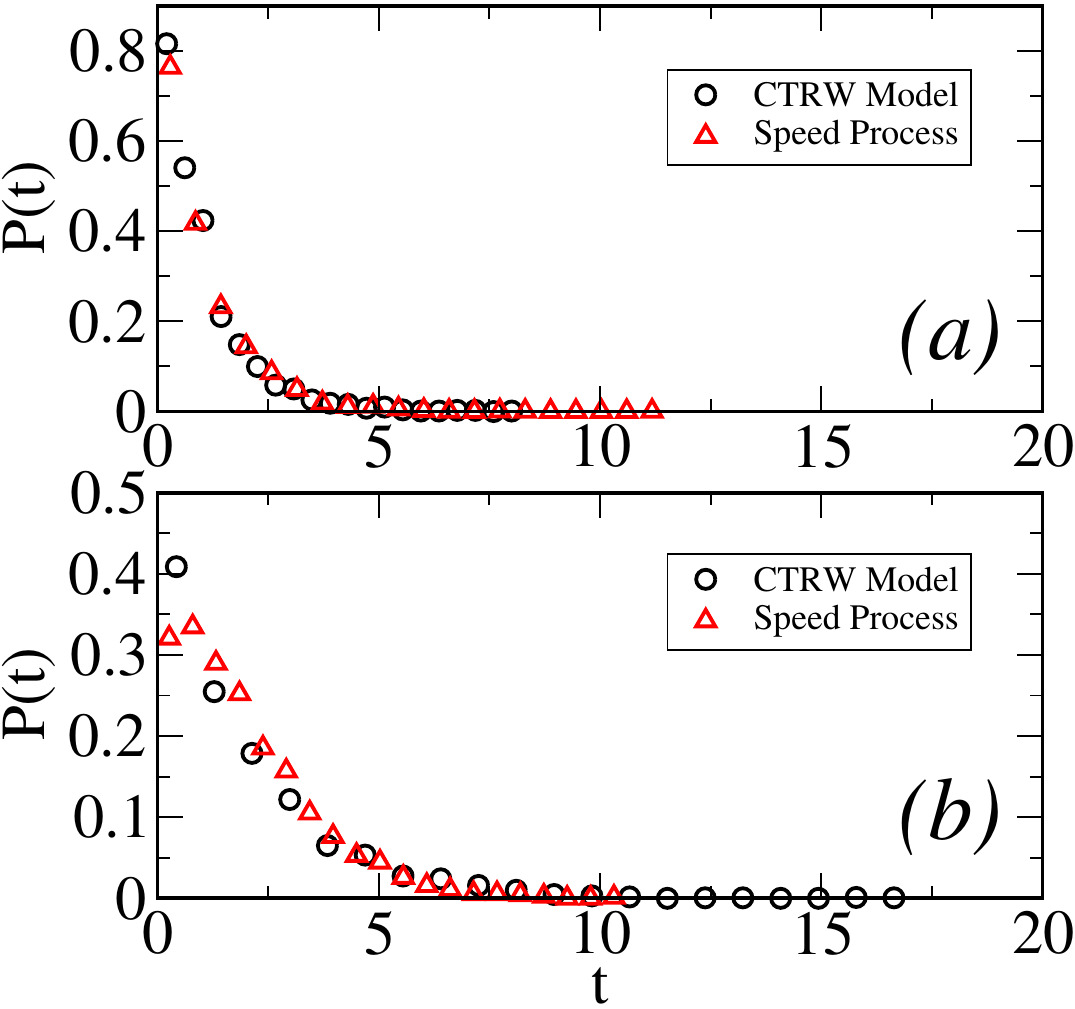}\caption{Waiting time distributions for trajectories with speeds (a) 1 and
(b) 0.2 in the CTRW model (black circle) and the TASEP Speed Process
(red circle). For speeds close to 1, the waiting time distribution
for the TASEP Speed Process matches closely with the exponential waiting
times of the CTRW model. However, for intermediate speeds, a clear
departure from exponential waiting time distribution is observed.}
\end{figure}

\section{Extensions of our approach}

This work can be extended to the case of a multi-species \emph{partially}
asymmetric exclusion process (ASEP). It was conjectured in {[}17{]}
that even in partially asymmetric case, the asymptotic velocity tends
to a uniformly distributed random variable. More precisely, in this
case, if we start with the initial conditions as before and look at
the motion of the second class particle, then:
\[
\lim_{t\rightarrow\infty}\frac{X(t)}{t}=U_{p}
\]
where $X(t)$ is the position of the second class particle at time
$t$ and $U_{p}$ is a uniform random variable between $[-(2p-1), (2p-1)]$
where $p$ is the rate of jumping to the right ($1-p$ being the rate
of jumping to the left). An Langevin description can be developed
for this as the evolution of density for first class particles is
given by:
\begin{equation}
\frac{\partial\rho(x,t)}{\partial t}+(2p-1)(1-2\rho(x,t))\frac{\partial\rho(x,t)}{\partial x}=0
\end{equation}
Even in the case of multi-species ASEP, our analysis goes through
and the fluctuations about the average velocity die out as $t^{-1/2}$. 

The finite lattice version of the multi-species problem $[21]$ offers
an interesting extension to the effective medium approach. The system
considered is a finite lattice with $n$ sites in which, each site
is occupied by a particle and its class is labeled by its initial
position on the lattice. The time evolution of the system is given
by the stochastic nearest neighbor exchange rule:

\[
ij\;\;\underrightarrow{\;rate\;1\;\;}\;\;ji\;\quad\text{for all }\:i<j
\]
If we wish to study the dynamics of a tagged particle of the $k$-th
class, it is clear that the problem is, again, reducible to a two
species problem with particles of $l$-th class ($l<k$) being equivalent
to first class particles, particles of $m$-th class ($m>k$) being
holes and the tagged particle being the second class particle. 

The motion of a tagged particle is strongly affected by the ends and
displays an interesting behaviour - initially, its dynamics of the
tagged particle mimics the dynamics of a tagged particle in TSP on
an infinite line. However, at later times, the particle reaches a
growing impenetrable region of density 1 (0) on the right (left) and
travels along with it, remaining at the moving end of this region
at subsequent times, till its absorbing position. This is expected
as the lattice is finite and after some time, clearly the first class
particles (holes) start to get accumulated at the right (left) boundary.
It is interesting to note that the absorbing position of a particle
whose initial position was $k$ is always $n-k$. This behaviour can
be described by CTRW model with jump rates given by the following
background density:

\begin{equation}
\begin{array}{cccc}
\rho(x,t)= & 0 & \textit{for } & x\leq l(t)\\
 & 1 & \textit{for } & x\geq n-r(t)\\
 & \frac{1}{2}(1-\frac{x-k}{t}) & \textit{for } & -t+l(t)<x-k<t-r(t)\\
 & 1 & \textit{for } & l(t)\leq x<k-t+l(t)\\
 & 0 & \textit{for } & k+t-r(t)\leq x<n-r(t)
\end{array}
\end{equation}

$l(t)$ and $r(t)$ are the mean widths of impenetrable regions of
density 1 and 0 on the right and left boundary respectively and they
satisfy:

\begin{equation}
\begin{array}{cccc}
r(t)= & 0 & for & t\leq(n-k)\\
 & t+(n-k)-2\sqrt{t(n-k)} & for & t_{s}\geq t>(n-k)\\
 & k & for & t>t_{s}\\
l(t)= & 0 & for & t\leq k\\
 & t+k-2\sqrt{tk} & for & t_{s}\geq t>k\\
 & n-k & for & t>t_{s}
\end{array}
\end{equation}

where $t_{s}$ is the mean time taken by the tagged particle to reach
its absorbing position and is given by $t_{s}=n+2\sqrt{k(n-k)}$.

\medskip{}

We simulated the process on a lattice of 1000 sites and looked at
the various trajectories of the particle labelled 200. Figure 6 shows
150 such trajectories and Figure 7 shows analogous results for the
effective medium description using a continuous time random walker
on a line with jump rates determined by the background density given
in Eq(13).

\begin{figure}[h]
\includegraphics[scale=0.3]{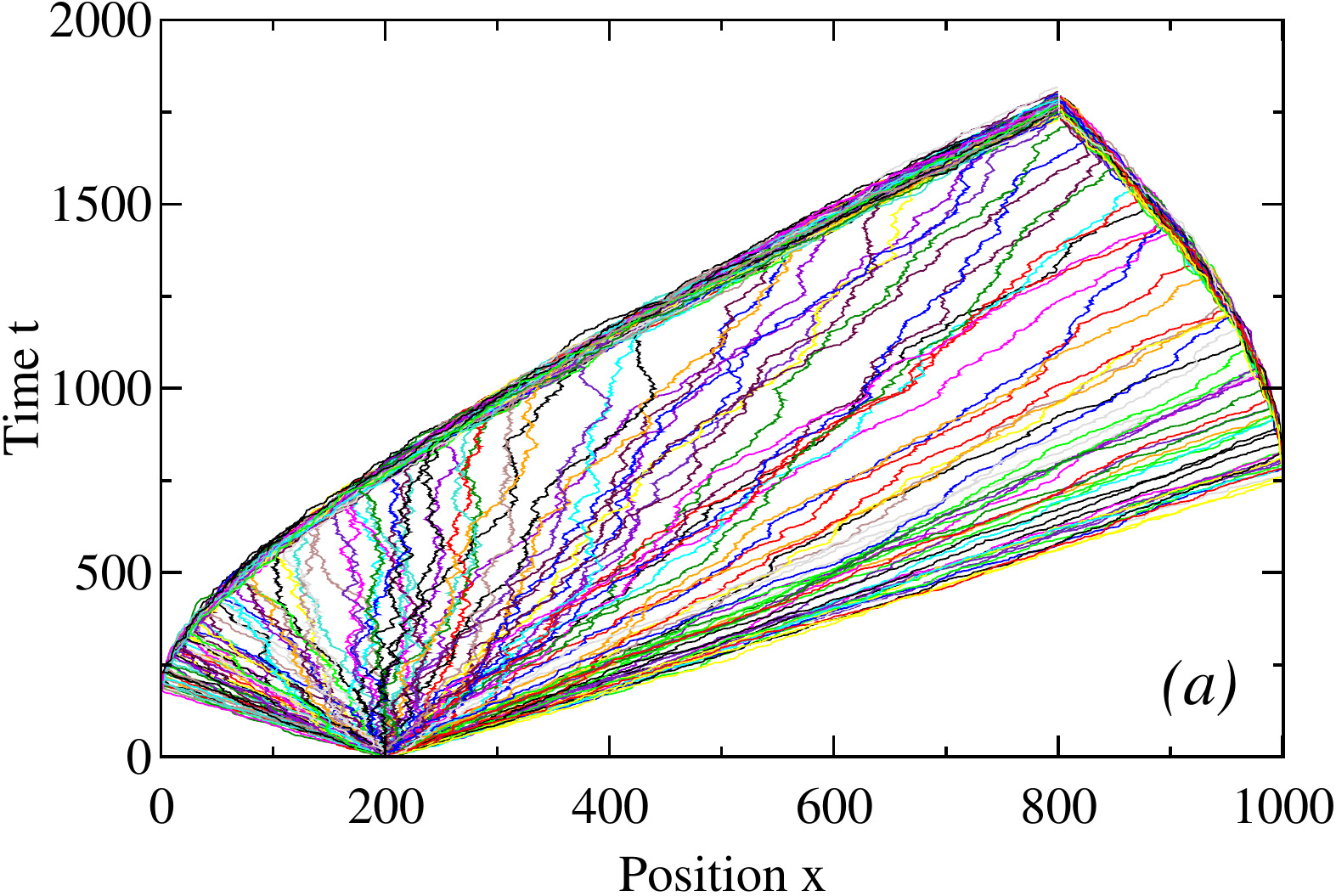}\caption{150 different trajectories of the particle labelled 200 in the finite
lattice version of TSP with 1000 particles. }
\end{figure}

\begin{figure}[h]
\includegraphics[scale=0.3]{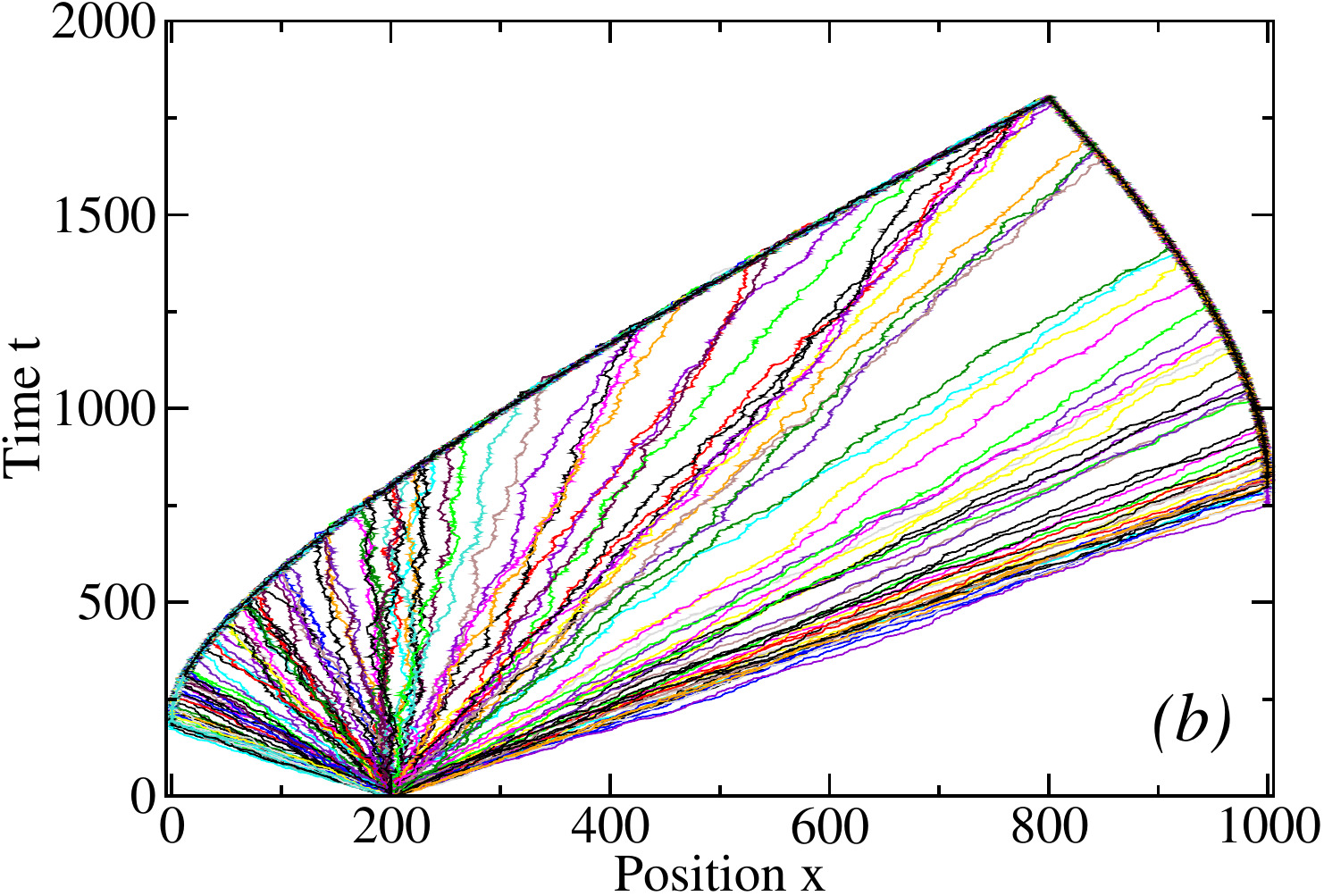}\caption{150 different trajectories of a continuous time random walker with
space time dependent jump rates determined by the evolving density
of first class particles defined by Eq(13). }
\end{figure}

\pagebreak{}

As an interesting variation to the multi-species problem, consider
a 1D lattice where each lattice site is occupied with a particle and
the class of each particle is labeled by its position on the lattice
with only the following nearest neighbor transitions allowed:

\[
ij\;\;\underrightarrow{\;rate\;(j-i)^{\alpha}\;\;}\;\;ji\;\quad\text{for all }\:i<j
\]
This model is clearly a better model for traffic flow if we visualize
the $x$-coordinate of particles to not be their position in real
space, but their relative order on the road as the overtakings between
two particles happen at rates proportional to the difference between
their labels (which is a proxy for the velocity). 

\begin{figure}[h]
\includegraphics[scale=0.28]{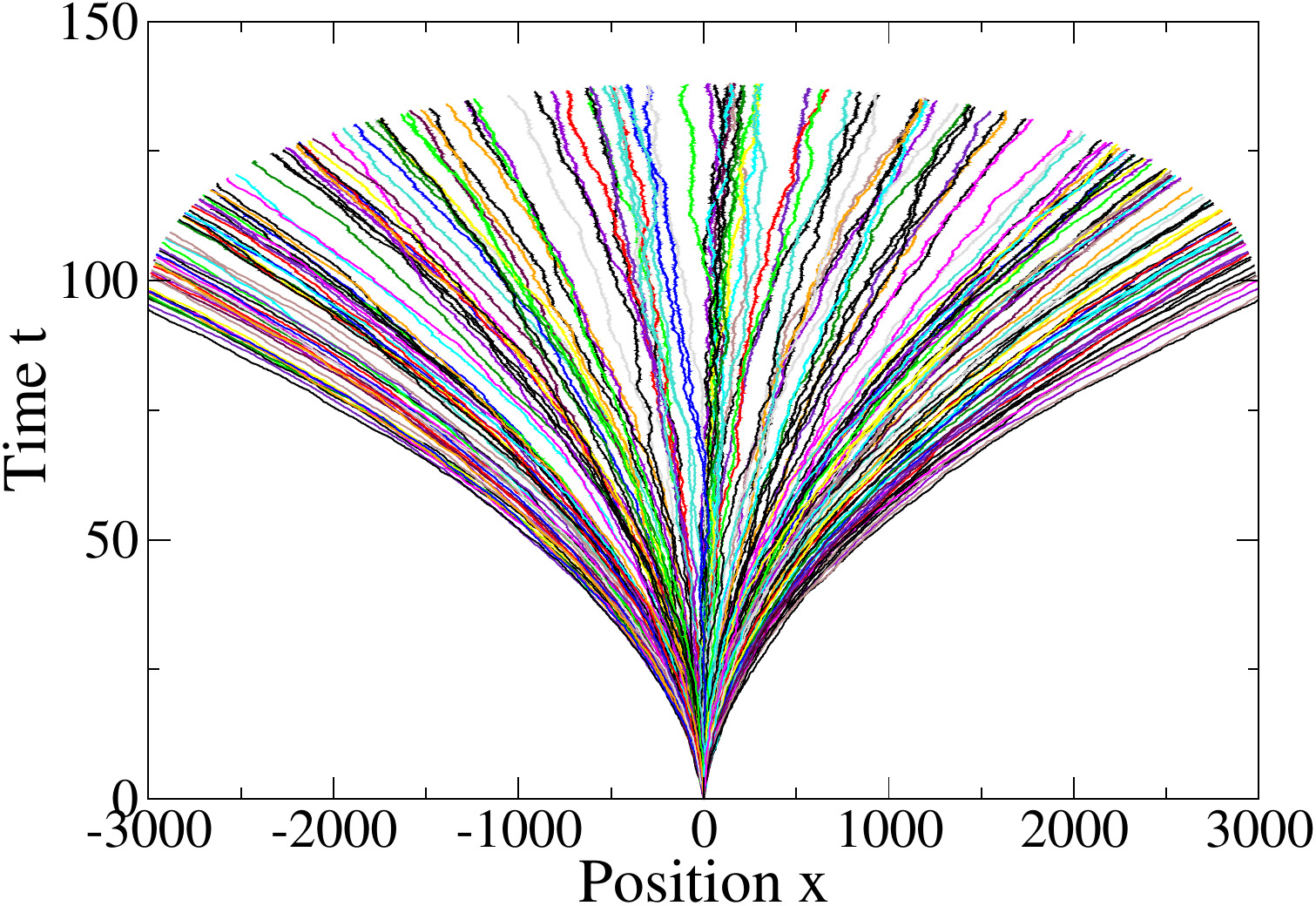}\caption{150 trajectories consisting of 3000 jumps made by a tagged particle
in the modified multi-species exclusion process with $\alpha=0.5$. }
\end{figure}

Some special cases of this model have been studied before $[6-8]$.
It is known that the steady state of such a model on a 1D lattice
with open boundary conditions and $\alpha=1$, in which there was
an additional feature that particles could enter the system from the
left end and leave from the right at rates depending on their labels,
can be obtained by a matrix product ansatz. This was later generalized
to obtain the steady state properties of this system on a ring. 

We consider the dynamics of a tagged particle in this modified multi-species
exclusion process on an infinite line for a general $\alpha$ where
particles do not enter or exit the system. This problem cannot be
reduced simply to the 2-species problems as each particle interacts
with every other particles differently. However, we find that something
analogous to the ``velocity selection'' in TSP happens in this process
as well. The trajectories of the tagged particle in this process seem
to belong to the family of curves $t=ax^{1-\alpha}$ for $\alpha\neq1$
and $t=\ln ax$ for $\alpha=1$ where $a$ parametrizes the trajectories.
A heuristic argument for this is as follows:if a particle has moved distance $x$ in time $t$, then the typical change in velocity it encounters with its neighbor is proportional to x. Then $dx/dt \sim x^{\alpha}$ implies that $ x \sim t^{1/(1-\alpha)}$.
for $\alpha\neq1$ and $t\thicksim\ln x$ for $\alpha=1$. 

\begin{figure}[h]
\includegraphics[scale=0.28]{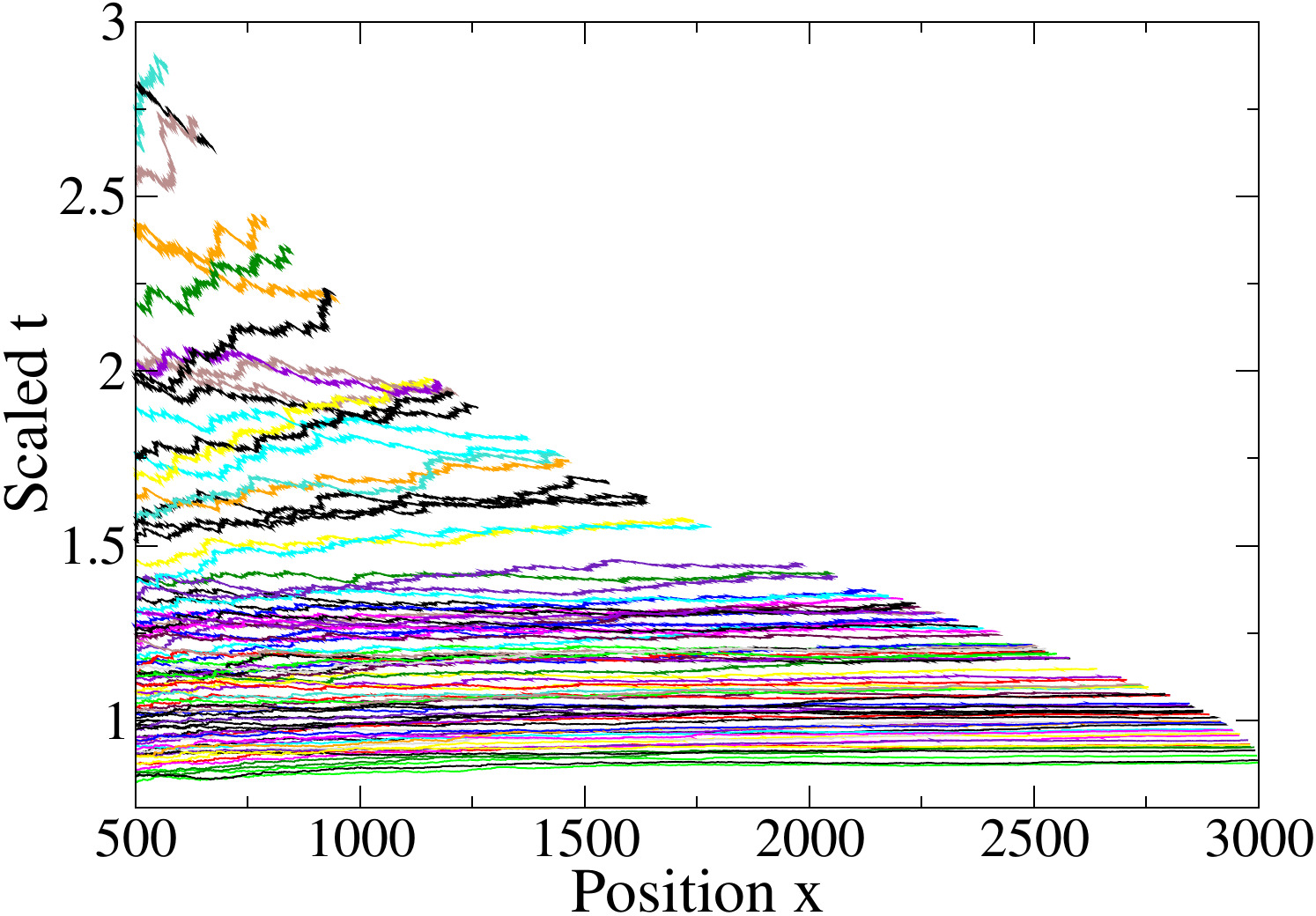}\caption{Trajectories of the tagged particle in the modified multi-species
exclusion process with $\alpha=0.5$ after time coordinate $t$ is
scaled as $\frac{t}{2\sqrt{x}}$. 150 different trajectories are plotted
showing that all the trajectories belong to a family of curves given
by $t\sim\sqrt{x}$. }
\end{figure}
We demonstrate numerical results of our scaling in Figures 8 and 9
for $\alpha=\frac{1}{2}$. Figure 8 shows 150 different space-time
trajectories of the process and Figure 9 shows the trajectories when
the time coordinate $t$ is scaled as $\frac{t}{2\sqrt{x}}$ where
we have only chosen the trajectories whose displacement always remains
positive. We see that $t/$$\sqrt{x}$ is nearly constant for each
trajectory, but different trajectories have very different values
of this variable. Finding the distribution of the asymptotic value
of $a$ over different trajectories remains an open problem.

\section{Summary and Concluding Remarks}

In summary, we discussed the effective medium approach to describe
the motion of a tagged particle in the time-evolving background other
particles. We provided a simple Langevin description of the dynamics,
that captures the key features of the large-scale behavior of TSP,
and also calculated the variance of the average velocity within one
history, and for different histories. We discussed how to improve
the effective medium description to take into account different additional
features of the trajectories. These were approximating the trajectory
as a biassed random walk, with rates of walk calculated from the steady
state of the TASEP, in the absence of the second-class particle. We
found that to get a quantitative agreement with the original process,
one has to incorporate the modification of the average density profile
that occurs near the second-class particle. Also, while the time evolution
of the original process was Markovian by definition, the evolution
of the projected process is non-Markovian. This is most easily seen
in the non-exponential distribution of waiting times between jumps
in the motion of the second-class particle. We proposed a non-Markovian
continuous time random walk with a distribution of waiting times between
jumps as a good description of this. We later extended our approach
to a finite lattice version of the TSP and studied the trajectories
of a tagged particle in this process. In this case, interesting end-effects
are seen which we explained using the effective medium approach. Lastly,
we looked at a modified multi-species exclusion process in which exchanges
between adjacent particles happened at rates proportional to the difference
in their labels. We showed that in this process too, the motion of
a tagged particle shows a behaviour in which it initially ``chooses''
a trajectory from a family of curves and sticks to it asymptotically.
A better understanding of our heuristic arguments, and numerical observations
about this process for a general $\alpha$ seems to be an interesting
problem for further study.

\end{document}